\documentclass[reprint, superscriptaddress]{revtex4-2}

\usepackage{tensor}
\usepackage{amsmath}
\usepackage{amssymb}
\usepackage{mathrsfs}
\usepackage{graphicx}
\usepackage{stmaryrd}
\usepackage{tikz-cd}
\usepackage{comment}
\usepackage[compat=1.1.0]{tikz-feynman}
\usepackage{twistor}
\usepackage{graphicx}
\usepackage{dcolumn}
\usepackage{bm}
\usepackage{hyperref}

\newcommand{\ba}{\mathbf{a}}
\newcommand{\m}{\mathrm{m}}

\renewcommand{\d}{\mathrm{d}}

\begin{document}


\title{MHV scattering of gluons and gravitons in chiral strong fields}

\author{Tim Adamo}
 \affiliation{School of Mathematics, University of Edinburgh, EH9 3FD, United Kingdom}
 \email{t.adamo@ed.ac.uk}
\author{Lionel Mason}
\author{Atul Sharma}
\affiliation{The Mathematical Institute, University of Oxford, OX2 6GG, United Kingdom}

\date{\today}

\begin{abstract}
We present all-multiplicity formulae for the tree-level scattering of gluons and gravitons in the maximal helicity violating (MHV) helicity configuration, calculated in certain chiral strong fields. The strong backgrounds we consider are self-dual plane waves in gauge theory and general relativity, which are treated exactly and admit a well-defined S-matrix. The gauge theory background-coupled MHV amplitude is simply a dressed analogue of the familiar Parke-Taylor formula, but the gravitational version has non-trivial new structures due to graviton tails. Both formulae have just one residual integral rather than the $n-2$ expected at $n$-points from space-time perturbation theory; this simplification arises from the integrability of self-dual backgrounds and their corresponding twistor description.  The resulting formulae pass several consistency checks and limit to the well-known expressions for MHV scattering of gluons and gravitons when the background becomes trivial.
\end{abstract}

\maketitle


\section{Introduction}

One hallmark of recent advances in the study of perturbative quantum field theories (QFTs) is the ability to find \emph{all-multiplicity} expressions for scattering amplitudes. A famous example is the Parke-Taylor formula for maximal helicity violating (MHV) tree-level colour-ordered amplitudes in four-dimensional Yang-Mills theory~\cite{Parke:1986gb}; this captures the semi-classical scattering of two negative helicity gluons with an arbitrary number of positive helicity gluons. Today there are all-multiplicity formulae for the full tree-level S-matrices of gauge theory, gravity and many other massless QFTs in diverse space-time dimensions (cf., \cite{Witten:2003nn,Roiban:2004yf,Cachazo:2012kg,Cachazo:2013hca,Cachazo:2014xea}), and a growing list of all-multiplicity formulae in some theories beyond tree-level (cf., \cite{ArkaniHamed:2010kv,Adamo:2013tsa,Geyer:2015jch,Bourjaily:2017wjl,Geyer:2018xwu,Bourjaily:2019gqu}). 

These formulae are in the `standard' setting of perturbative QFT -- where the perturbative expansion occurs around a trivial field configuration -- but there are myriad settings of physical interest where scattering (or its analogues) in the presence of non-trivial, or \emph{strong}, background fields must be considered. Just a few examples are: QED processes in high-intensity lasers~\cite{DiPiazza:2011tq,King:2015tba}; QCD processes in the vicinity of heavy ion collisions~\cite{Gelis:2010nm}; the computation of $n$-spectra in slow roll inflation~\cite{Mukhanov:1990me,Maldacena:2002vr}; and strongly coupled conformal field theory correlators via the AdS/CFT correspondence~\cite{Aharony:1999ti}. In these settings, the strong fields are treated as an exact classical background around which QFT is operationalised via background perturbation theory (cf., \cite{Abbott:1981ke}).

There is a conspicuous gap between the all-multiplicity formulae for scattering on trivial backgrounds and what is known about the S-matrix (or its analogues) in strong backgrounds. Indeed, the best results at tree-level in strong backgrounds are at 4-points: QED Compton or trident scattering in an electromagnetic plane wave (e.g., \cite{DiPiazza:2011tq,King:2015tba}); some 4-point boundary correlation functions in anti-de Sitter space~\cite{DHoker:1999kzh,Raju:2012zs}; or gluon scattering on a Yang-Mills plane wave background~\cite{Adamo:2018mpq}. While Feynman rules on strong backgrounds can often be written explicitly, performing computations with them is difficult beyond the very lowest orders in perturbation theory. 

However, all-multiplicity formulae on trivial backgrounds have nothing to do with the traditional Feynman expansion: instead, they are derived from alternative formulations of field theory (e.g., as a worldsheet theory~\cite{Witten:2003nn,Mason:2013sva}) or from considerations of locality, unitarity and gauge invariance (e.g., \cite{Britto:2005fq,Arkani-Hamed:2016rak,Rodina:2018pcb}). Yet on strong backgrounds, there are apparent obstructions to applying these innovative techniques. For instance, functional degrees of freedom in the background mean that the singularity structure of amplitudes is no longer described simply by rational functions. 

With this in mind, it is natural to ask: is it possible to find all-multiplicity formulae for scattering in strong backgrounds? Here, we answer this question in the affirmative for tree-level MHV scattering on a particularly simple class of gauge and gravitational backgrounds in four space-time dimensions. The resulting formulae are the first all-multiplicity expressions in \emph{any} strong background (besides those which vanish trivially or result from approximations~\cite{DiPiazza:2010mv}) and are structurally reminiscent of all-multiplicity formulae in a trivial background, but with significant novel features. 


\section{The chiral backgrounds}

The strong backgrounds we consider are \emph{self-dual plane waves} (SDPWs) in four-dimensions. These are special cases of plane waves whose field strengths are self-dual. The physical interpretation of such backgrounds is as a coherent superposition of positive helicity gluons or gravitons. A plane wave is a solution to the vacuum equations of motion with special symmetry properties: it has 5 symmetries whose generators commute to form a Heisenberg algebra with centre given by a covariantly constant null symmetry~\cite{Baldwin:1926,Ehlers:1962zz,Trautman:1980bj,Adamo:2017nia}.

In four-dimensions, it will be convenient to work with 2-spinor notation, trading tensor indices for a pair of SL$(2,\C)$ Weyl spinor indices of opposite helicity (cf., \cite{Penrose:1984uia}). The (complexified) Minkowski line element in light-front coordinates is
\be\label{Mink}
\d s^2=2\left(\d x^{+}\,\d x^{-}-\d z\,\d\tilde{z}\right)\,,
\ee
where the transverse coordinate $\tilde{z}$ is not required to be the complex conjugate of $z$. These coordinates are neatly packaged into a $2\times 2$ matrix:
\be\label{2sp1}
x^{\alpha\dot\alpha}=\left(\begin{array}{c c} x^+ & \tilde{z} \\
                                                                   z & x^- \end{array}\right)\,,
\ee
where $(\alpha,\dot{\alpha})$ are SL$(2,\C)$ Weyl spinor indices of negative and positive chirality, respectively. The line element \eqref{Mink} in 2-spinor notation is $\d s^2=\d x^{\alpha\dot\alpha}\,\d x_{\alpha\dot\alpha}$, with spinor indices raised and lowered using the two-dimensional Levi-Civita symbols $\epsilon_{\alpha\beta}$, $\epsilon_{\dot\alpha\dot\beta}$, etc.

Let $F_{ab}$ be the field strength of a gauge potential. This admits a spinor decomposition
\be\label{Fdecomp}
F_{\alpha\dot\alpha\beta\dot\beta}=\epsilon_{\alpha\beta}\,\tilde{F}_{\dot\alpha\dot\beta}+\epsilon_{\dot\alpha\dot\beta}\,F_{\alpha\beta}\,,
\ee
where $\tilde{F}_{\dot\alpha\dot\beta}$ and $F_{\alpha\beta}$ are the self-dual (SD) and anti-self-dual (ASD) components of the field strength, respectively. They are symmetric in their spinor indices and valued in the adjoint representation of the gauge group. The self-duality condition is simply that $F_{\alpha\beta}=0$; non-trivial solutions (i.e., $\tilde{F}_{\dot\alpha\dot\beta}\neq0$) in Lorentzian signature are only possible for complex gauge fields, which is why we allow $\tilde{z}\neq\bar{z}$.

Combining the self-duality and plane wave conditions forces the gauge field to be valued in the Cartan of the gauge group~\cite{Trautman:1980bj,Adamo:2017nia}. The gauge potential and field strength are
\be\label{YMSD1}
A=-f(x^-)\,\d\tilde z\,, \quad \tilde{F}_{\dot\alpha\dot\beta}=\dot{f}(x^-)\,\tilde{\iota}_{\dot\alpha}\tilde{\iota}_{\dot\beta}\,,\quad F_{\alpha\beta}=0\,,
\ee
where $f(x^-)$ is Cartan-valued and $\dot{f}\equiv \partial_- f$. The covariantly constant null symmetry associated with the background is generated by the vector 
\be\label{null}
n^{\alpha\dot\alpha}=\iota^{\alpha}\,\tilde{\iota}^{\dot\alpha}\,, \qquad \iota^{\alpha}=\left(\begin{array}{c} 1 \\ 0 \end{array}\right)=\tilde{\iota}^{\dot\alpha}\,.
\ee 
It is also useful to define the constant spinors $o_{\alpha},\tilde o_{\dot\alpha}$ forming spinor dyads with $\iota^{\alpha}$ and $\tilde\iota^{\dot\alpha}$: $\iota^{\alpha}\,o_{\alpha}=1=\tilde\iota^{\dot\alpha}\,\tilde o_{\dot\alpha}$.

It is easy to see that \eqref{YMSD1} solves the Yang-Mills equations for \emph{any} Cartan-valued $f$, but we restrict our attention to `sandwich' self-dual plane waves for which $\dot{f}$ is compactly supported in $x^-$; such profiles admit a well-defined gluon S-matrix~\cite{Schwinger:1951nm,Adamo:2017nia}.

In the gravitational context, self-duality is a condition on the Weyl curvature tensor, which admits a spinor decomposition:
\be\label{Wdecomp}
C_{\alpha\dot\alpha\beta\dot\beta\gamma\dot\gamma\delta\dot\delta}=\epsilon_{\alpha\beta}\,\epsilon_{\gamma\delta}\,\tilde{\Psi}_{\dot\alpha\dot\beta\dot\gamma\dot\delta}+\epsilon_{\dot\alpha\dot\beta}\,\epsilon_{\dot\gamma\dot\delta}\,\Psi_{\alpha\beta\gamma\delta}\,,
\ee
where $\tilde{\Psi}_{\dot\alpha\dot\beta\dot\gamma\dot\delta}$ and $\Psi_{\alpha\beta\gamma\delta}$ are the SD and ASD parts of the Weyl tensor. Self-duality gives $\Psi_{\alpha\beta\gamma\delta}=0$ and, combined with the plane wave conditions, one finds the metric (in Einstein-Rosen coordinates):
\be\label{GRSD1}
\d s^2 = 2\left(\d x^{+}\,\d x^{-}-\d z\,\d\tilde{z}+f(x^{-})\,\d \tilde{z}^2\right)\,.
\ee
We have abused notation to denote the functional freedom by $f(x^-)$; it should be clear from the context whether $f$ is Cartan valued (gauge theory) or a pure function (gravity). The SDPW metric \eqref{GRSD1} solves the vacuum Einstein equations for any $f$, with Weyl curvature
\be\label{GRSD2}
\tilde{\Psi}_{\dot\alpha\dot\beta\dot\gamma\dot\delta}=-\ddot{f}(x^-)\,\tilde{\iota}_{\dot\alpha}\tilde{\iota}_{\dot\beta}\tilde{\iota}_{\dot\gamma}\tilde{\iota}_{\dot\delta}\,, \qquad \Psi_{\alpha\beta\gamma\delta}=0\,.
\ee
The sandwich condition for a SDPW space-time is that $\ddot{f}$ be compactly supported in $x^-$; this ensures existence of a well-defined tree-level graviton S-matrix~\cite{Gibbons:1975jb,Adamo:2017nia}.


\section{SD plane wave Kinematics}

Strong plane wave backgrounds admit momentum eigenstate representations for free field perturbations~\cite{Wolkow:1935zz,Friedlander:2010eqa,Ward:1987ws} and the Feynman rules for gluons and gravitons can be explicitly constructed from these in any dimension~\cite{Adamo:2017nia,Adamo:2018mpq}. 

\paragraph{Yang-Mills} A gluon on a Yang-Mills SDPW can be reduced to a scalar field of charge $e$ with respect to the Cartan-valued background. This is given by $\e^{\im\phi_K}$ where $\phi_K$ also solves the Hamilton-Jacobi equation. Thus $$(\partial_a-\im\,e\,A_a)\e^{\im\phi_K}=\im\, K_a\,\e^{\im\phi_K}$$ defines a null vector field $K_a$ on the background. In 4-dimensions, the spinor-helicity formalism (cf., \cite{Dixon:2013uaa}) can be used to further refine the on-shell kinematics of gluon and graviton perturbations in plane waves~\cite{Adamo:2019zmk}. For SDPW backgrounds, these on-shell kinematics are themselves \emph{chiral}. The on-shell 4-momentum of a gluon moving through a SDPW gauge field background admits the 2-spinor decomposition:
\be\label{glmom}
K_{\alpha\dot\alpha}(x^-)=\lambda_{\alpha}\,\tilde{\Lambda}_{\dot\alpha}(x^-)\,, \quad \tilde{\Lambda}_{\dot\alpha}=\tilde{\lambda}_{\dot\alpha}+\frac{e\, f(x^-)}{\la \iota\,\lambda\ra}\,\tilde{\iota}_{\dot\alpha}\,,
\ee
where 
$\la a\,b\ra=a^{\alpha}\,b_{\alpha}=\epsilon^{\alpha\beta}a_{\beta}b_{\alpha}$, $[\tilde{a}\,\tilde{b}]=\tilde{a}^{\dot\alpha}\,\tilde{b}_{\dot\alpha}=\epsilon^{\dot\alpha\dot\beta}\tilde{a}_{\dot\beta}\tilde{b}_{\dot\alpha}$.

The constant null vector $k_{\alpha\dot\alpha}=\lambda_{\alpha}\tilde{\lambda}_{\dot\alpha}$ is the gluon's momentum \emph{before} it enters the SDPW background. As the gluon traverses the background, the momentum spinor $\tilde{\lambda}_{\dot\alpha}$ is `dressed': it picks up non-trivial dependence on lightfront time $x^-$ through the background wave profile. Chirality of the SDPW background ensures that only the positive chirality momentum spinor is dressed: the negative chirality spinor $\lambda_{\alpha}$ is un-changed by the background.

This chirality is also present at the level of positive/negative helicity gluon polarizations, which can be written in lightcone-transverse gauge as:
\be\label{glpol}
\cE^{(+)}_{\alpha\dot\alpha}=\frac{\iota_{\alpha}\,\tilde{\Lambda}_{\dot\alpha}}{\la\iota\,\lambda\ra}\,, \qquad \cE^{(-)}_{\alpha\dot\alpha}=\frac{\lambda_{\alpha}\,\tilde{\iota}_{\dot\alpha}}{[\tilde{\iota}\,\tilde{\lambda}]}\,.
\ee
These obey $n\cdot\cE^{(\pm)}=0=K\cdot\cE^{(\pm)}$, and only the positive helicity polarization is dressed by the background, through $\tilde{\Lambda}_{\dot\alpha}(x^-)$.  

\paragraph{Gravity} For a graviton propagating on a SDPW space-time, the kinematics and polarizations are also chiral. The on-shell graviton 4-momentum is
\be\label{grmom}
K_{\alpha\dot\alpha}(x^-)=\lambda_{\alpha}\,\tilde{\Lambda}_{\dot\alpha}(x^-)\,, \quad \tilde{\Lambda}_{\dot\alpha}=\tilde{\lambda}_{\dot\alpha}-\frac{\la o\,\lambda\ra\,[\tilde{\iota}\,\tilde{\lambda}]\,f(x^-)}{\la \iota\,\lambda\ra}\,\tilde{\iota}_{\dot\alpha}\,,
\ee
where $k_{\alpha\dot\alpha}=\lambda_{\alpha}\tilde{\lambda}_{\dot\alpha}$ is the graviton momentum before entering the background. Positive and negative helicity graviton polarizations (in transverse, traceless lightcone gauge) are
\be\label{grpol1}
\begin{split}
\cE^{(+)}_{\alpha\dot\alpha\beta\dot\beta}&=\frac{\iota_{\alpha}\,\iota_{\beta}}{\la\iota\,\lambda\ra^2}\left(\tilde{\Lambda}_{\dot\alpha}\,\tilde{\Lambda}_{\dot\beta}-\im\,\dot{f}\,\tilde{\iota}_{\dot\alpha}\tilde{\iota}_{\dot\beta}\right)\,,\\
\cE^{(-)}_{\alpha\dot\alpha\beta\dot\beta}&=\frac{\lambda_{\alpha}\,\lambda_{\beta}\,\tilde{\iota}_{\dot\alpha}\tilde{\iota}_{\dot\beta}}{[\tilde{\iota}\,\tilde{\lambda}]^2}\,.
\end{split}
\ee
Once again, only the positive helicity configuration is dressed by the background but there is an additional tail term with coefficient $\dot f$ in \eqref{grpol1}.


\section{MHV amplitudes}

It is straightforward to prove that all tree-level amplitudes in SDPWs with one or zero negative helicity gluons or gravitons vanish; this is a consequence of the integrability of the self-dual sector in gauge theory and gravity \cite{Mason:2008jy}. Intuitively, this is not surprising: such amplitudes vanish on a trivial background, and one can imagine the SDPW constructed from a coherent superposition of further positive helicity gluons/gravitons. Hence, the first non-trivial tree amplitudes in SDPW backgrounds should appear for the MHV helicity configuration: two negative helicity gluons/gravitons and an arbitrary number of positive helicity gluons/gravitons.

Some basic features of tree-amplitudes in strong plane waves can be deduced from general properties of perturbation theory for these backgrounds. One is the absence of total 4-momentum conservation: the free function (or Cartan-valued function) defining the background breaks Poincar\'e invariance in the $x^-$-direction. Momentum conservation in the $x^-$-direction is replaced by lightfront integrals that cannot be performed explicitly due to the free function in the background. In the $x^+$, $z$ and $\tilde{z}$-directions, momentum is still conserved, so there will be a factor of
\be\label{3mom}
\delta^{3}_{+,\perp}\!\left(\sum_{i=1}^{n}k_i\right)\,,
\ee
in all amplitudes. These surviving components of momentum conservation can be expressed as:
\be\label{3mom1}
\sum_{i=1}^{n}K_{i\alpha\dot\alpha}(x^-)=\sum_{i=1}^{n}K_{i-}(x^-)\,\iota_{\alpha}\,\tilde{\iota}_{\dot\alpha}\,,
\ee
where $K_{-}$ is the component of the dressed momentum in the $x^-$-direction. In the gauge theory setting we also have overall charge conservation $\sum_i e_i=0$ with respect to the Cartan-valued background.


Naively, there should be one residual lightfront integral for each interaction vertex contributing to a generic Feynman diagram from the background field Lagrangian. Since both Yang-Mills and gravity behave like cubic theories (cf., \cite{Bern:2019prr}), this leads to an expected $n-2$ residual lightfront integrals for $n$-point tree amplitudes.  However, for SDPW MHV amplitudes, a remarkable simplification expresses them in terms of a single $x^-$-integral that replaces the last component of momentum conservation.

\subsection{Gluon amplitude}

We take our particles' colours to have well-defined charges $e_i$ with respect to the Cartan background. The partial amplitude with colour-ordering $\{1,2,\ldots,n\}$ for MHV gluon scattering on a SDPW gauge background is then given by:
\be\label{glMHV1}
\frac{\la i\,j\ra^{4}}{\la 1\,2\ra\,\la2\,3\ra\cdots\la n-1\,n\ra\,\la n\,1\ra}\,\int_{-\infty}^{+\infty}\d x^{-}\,\e^{\im\,\cF_{n}(x^{-})}\,,
\ee
where the overall momentum (and charge) conserving delta functions \eqref{3mom} have been omitted, gluons $i$ and $j$ are negative helicity and the remainder are positive helicity. The exponential phase appearing in the residual lightfront integration is a $n$-point generalization of an object appearing at lower points in strong field QED, known as the \emph{Volkov exponent}~\cite{Wolkow:1935zz,Seipt:2017ckc}. This is defined by a composite momentum $\mathbb{K}$ built from any set of $n-1$ distinct dressed gluon momenta; for instance,
\be\label{glVE0}
\mathbb{K}^{\alpha\dot\alpha}(x^{-}):=\sum_{i=1}^{n-1}K_{i}^{\alpha\dot\alpha}(x^-)\,.
\ee
The Volkov exponent is then given by
\be\label{glVE}
 \cF_n(x^-):= \int^{x^-}\!\!\!\,\frac{\d s\;\mathbb{K}^{2}(s)}{2\,\la\iota|\mathbb{K}(s)|\tilde{\iota}]} =\sum_{i=1}^{n}\int^{x^-}\!\!\!\d s\,
 K_{i-}(s) 
 \,,
\ee
the equality following by 3-momentum conservation \eqref{3mom1} and the on-shell condition $K_n^2 = 0$.

The ratio of angle brackets appearing in \eqref{glMHV1} matches the Parke-Taylor formula on a trivial background~\cite{Parke:1986gb}, since undotted momentum spinors are not dressed in a SDPW background. However, it is surprising that there is only a \emph{single} residual lightfront integral, regardless of the number of positive helicity gluons in the amplitude. In a trivial background, a non-local field redefinition enables the Yang-Mills Lagrangian to be recast such that all MHV amplitudes arise from a single interaction vertex~\cite{Cachazo:2004kj,Mason:2005zm,Mansfield:2005yd}. Remarkably, this argument also holds for the background Lagrangian on a SDPW, explaining the simplicity of the residual lightfront integral~\cite{Adamo:2020tba}. 

We have checked this formula against traditional Feynman diagram computations in background perturbation theory at 3- and 4-points, and it is also easy to show that the formula has the correct trivial background limit. In this case, the Volkov exponent and residual lightfront integral reduce to momentum conservation in the $x^-$-direction.

Beyond these somewhat piecemeal checks, \eqref{glMHV1} can actually be \emph{derived} from the space-time generating functional for MHV amplitudes. This generating functional can be lifted to twistor space, where integrability of the SD sector is manifest, enabling an all-multiplicity expansion \cite{Mason:2008jy}. This derivation extends to SDPW backgrounds and will be presented in~\cite{Adamo:2020tba}.

\subsection{Graviton amplitude}

On a trivial background, the tree-level graviton MHV amplitude depends on momentum spinors of both chiralities~\cite{Berends:1988zp}, and has a compact structure as a once-reduced $(n-2)\times(n-2)$ determinant discovered by Hodges~\cite{Hodges:2012ym}. The graviton MHV amplitude in a SDPW space-time requires more than a simple dressing -- as in \eqref{glMHV1} for Yang-Mills -- and is significantly more complicated. Nevertheless, this added complication has a simple physical interpretation and the final result is remarkably simple considering the apparent complexity of perturbative gravity arising from the Einstein-Hilbert Lagrangian. We find a sum over contributions from once-reduced $(n+t-2)\times(n+t-2)$ determinants where $t$ is the number of interactions with tails in the gravitational waves arising from previous interactions.

The formula involves a summation over partitions of the set of positive helicity external gravitons. For the MHV configuration, take gravitons 1 and 2 to be negative helicity and gravitons $\{3,\ldots,n\}$ to be positive helicity. Let $\ba_{\m}$ for $\m=1,\ldots,t$ be $t$ disjoint subsets of $\{3,\ldots,n\}$ that obey $|\mathbf{a}_{\mathrm{m}}|\geq2$ and $|\ba_1|+\cdots+|\ba_t|:=|\ba|\leq n-3$, so that $t$ is a non-negative integer with $t\leq\lfloor\frac{n-3}{2}\rfloor$. We denote the complement by:
\be\label{part2}
\bar{\mathbf{a}}:=\{3,\ldots,n\}\setminus\bigcup_{\mathrm{m}=1}^{t}\mathbf{a}_\mathrm{m}\,.
\ee
The MHV amplitude involves a sum over $t$, and further over all possible partitions $\ba_{1},\ldots,\ba_{t}$ for each $t$.

Given $t$ and  $\ba_{1},\ldots,\ba_{t}$,  define an $(n+t-2)\times(n+t-2)$ matrix with block decomposition:
\be\label{gLap}
\cH[\ba_1,\ldots,\ba_t]=\left(\begin{array}{c c c}
																\mathbb{H} & \mathfrak{h} &  0 \\
																0 & \mathbb{D} & \tau \\
																\mathfrak{c} & \mathfrak{t} & \T
																\end{array}\right)\,.
\ee
The $(n-|\ba|-2)\times(n-|\ba|-2)$ block $\HH$ is given by
\begin{equation*}
\HH_{ij}=\frac{[\tilde{\Lambda}_i\,\tilde{\Lambda}_{j}]}{\la i\,j\ra}(x^-)\,, \quad \HH_{ii}=-\sum_{j\neq i}\HH_{ij}(x^{-})\,\frac{\la 1\,j\ra\,\la2\,j\ra}{\la1\,i\ra\,\la2\,i\ra}\,,
\end{equation*}
with indices $i,j\in\bar{\ba}$; this is a dressed analogue of the matrices appearing in \cite{Hodges:2012ym}. The $|\ba|\times|\ba|$ block $\mathbb{D}$ and $t\times t$ block $\T$ are diagonal:
\begin{align*}
\mathbb{D}&= \mathrm{diag}\left(-\frac{[\tilde{\iota}\,j]\,\la1\,\iota\ra\,\la2\,\iota\ra}{\la\iota\,j\ra\,\la1\,j\ra\,\la2\,j\ra}\right)_{j\in\cup_{\m}\mathbf{a}_{\mathrm{m}}}, \\
\T &= \mathrm{diag}\left(-\sum_{j\notin\mathbf{a}_{\mathrm{m}}}\frac{[\tilde{\iota}\,j]\,\la1\,j\ra\,\la2\,j\ra}{\la\iota\,j\ra\,\la1\,\iota\ra\,\la2\,\iota\ra}\right)_{\m=1,\ldots,t}.
\end{align*}
For the off-diagonal blocks, $\mathfrak{h}$ is $(n-|\ba|-2)\times|\ba|$ and $\mathfrak{c}$ is $t\times(n-|\ba|-2)$:
\begin{equation*}
\mathfrak{h}_{ij}=\frac{[\tilde{\Lambda}_i\,\tilde{\Lambda}_{j}]}{\la i\,j\ra}(x^-)\,, \quad \mathfrak{c}_{\m i}=\frac{[\tilde{\iota}\,i]}{\la\iota\,i\ra}\,,
\end{equation*}
where $i\in\bar{\ba}$ and $j\in\cup_{\m}\ba_{\m}$. Finally, $\tau$ is $|\ba|\times t$ and $\mathfrak{t}$ is $t\times |\ba|$:
\begin{equation*}
\tau_{j\m}=\left\{\begin{array}{l}
				\frac{[\tilde{\iota}\,j]}{\la\iota\,j\ra} \:\mbox{ if } j\in\ba_{\m} \\
				0 \:\mbox{ otherwise}
				\end{array}\right.\,, \quad
\mathfrak{t}_{\m j}=\left\{\begin{array}{l}
						 \frac{[\tilde{\iota}\,j]}{\la\iota\,j\ra} \:\mbox{ if } j\notin\ba_{\m} \\
				0 \:\mbox{ otherwise}
				\end{array}\right.\,,
\end{equation*} 
for $j\in\cup_{\m}\ba_\m$ and $\m=1,\ldots,t$. These blocks emerge through an application of the matrix tree theorem to a space-time generating functional of graviton MHV amplitudes, which is also the reason for the vanishing of two of the off-diagonal blocks in \eqref{gLap}.

With these ingredients, the $n$-point tree-level graviton MHV amplitude on a SDPW background is:
\begin{multline}\label{grMHV}
\sum_{t=0}^{\lfloor\frac{n-3}{2}\rfloor} \int\limits_{-\infty}^{+\infty}\d x^{-} \sum_{\substack{i\in\bar{\mathbf{a}} \\\mathbf{a}_1,\ldots,\mathbf{a}_{t}}}\left|\mathcal{H}^{i}_{i}[\mathbf{a}_{1}\cdots\mathbf{a}_t]\right| \frac{\la1\,2\ra^{5}\,[\tilde{\iota}\,i]}{\la i\,1\ra\,\la i\,2\ra^2\,[\tilde{\iota}\,2]} \\
\times\,\im^{t-|\ba|}\,\e^{\im\,F_n}\,\prod_{\mathrm{m}=1}^{t} f^{(|\ba_{\mathrm{m}}|-1)}\,.
\end{multline}
The notation $|\cH^{i}_{i}|$ denotes the determinant of $\cH$ with row and column $i$ removed; $f^{(d)}:=\partial^{d}_{-}f$ for $f(x^-)$ the function defining the SDPW background; and $F_n$ is the gravitational Volkov exponent
\be\label{grVE}
F_n(x^-):=\int^{x^-}\d s\,\frac{g^{ab}(s)\,\mathbb{K}_{a}(s)\,\mathbb{K}_{b}(s)}{2\,\la\iota|\mathbb{K}(s)|\tilde{\iota}]}\,,
\ee
where $g_{ab}$ is the SDPW metric \eqref{GRSD1} and $\mathbb{K}_{a}$ is defined by any set of $n-1$ graviton momenta, as in \eqref{glVE0}.

Again, there is only a single residual lightfront integral at arbitrary multiplicity, but unlike the gluon amplitude, the integrand includes substantially more than the Volkov exponent: $\cH$ and $f^{(d)}$ introduce additional dependence on $x^-$. Terms in \eqref{grMHV} with $t\geq1$ are \emph{tail} contributions to the amplitude, arising due to the failure of Huygens' principle for perturbative gravity on a plane wave space-time~\cite{Harte:2013dba}. These tail terms represent contributions from background gravitons to the scattering, through insertions of (derivatives of) the background wave profile $f$. 

The formula \eqref{grMHV} has the correct little group weights in each external graviton, is permutation symmetric in the positive and negative helicity gravitons (the latter follows from the underlying MHV generating functional, but we have checked it explicitly through 6-points), and has been checked against Feynman diagram calculations in background perturbation theory at 3- and 4-points. In the flat background limit only the $t=0$ term survives and $|\cH^{i}_{i}|\rightarrow|\HH^{12i}_{12i}|$, where $\HH$ is now the Hodges matrix on a flat background~\cite{Hodges:2012ym}. The residual lightfront integral of the Volkov exponent completes \eqref{3mom} to 4-momentum conservation, and the remaining kinematic contribution to the amplitude is:
\begin{multline}
\sum_{i=3}^{n}\left|\HH^{12i}_{12i}\right| \frac{\la1\,2\ra^{5}\,[\tilde{\iota}\,i]}{\la i\,1\ra\,\la i\,2\ra^2\,[\tilde{\iota}\,2]}=\frac{|\HH^{123}_{123}|\,\la1\,2\ra^{6}}{\la1\,3\ra^2\,\la2\,3\ra^{2}} \sum_{i=3}^{n}\frac{\la i\,1\ra\,[\tilde{\iota}\,i]}{\la1\,2\ra\,[\tilde{\iota}\,2]} \\
=-\frac{|\HH^{123}_{123}|\,\la1\,2\ra^{6}}{\la1\,3\ra^2\,\la2\,3\ra^{2}} \sum_{i=3}^{n}\frac{\la1|k_i|\tilde{\iota}]}{\la1|k_2|\tilde{\iota}]}= \frac{|\HH^{123}_{123}|\,\la1\,2\ra^{6}}{\la1\,3\ra^2\,\la2\,3\ra^{2}}\,,
\end{multline}
with the various equalities following from basic properties of $\HH$ and 4-momentum conservation. This leaves precisely Hodges' formula for the gravitational MHV amplitude on a flat background~\cite{Hodges:2012ym}, in a representation agreeing with previous expressions~\cite{Bern:1998sv,Nguyen:2009jk}. 
 
The formula \eqref{grMHV} can be derived directly from a space-time generating functional for all MHV amplitudes in general relativity~\cite{Mason:2008jy}, expanded using twistor theory to manifest the integrability of the SD sector as in \cite{Adamo:2013tja}. Details of this (which includes a novel derivation of the Hodges formula in a flat background) will appear in~\cite{Adamo:2020tba}.


\section{Discussion}

The formulae \eqref{glMHV1} and \eqref{grMHV} constitute the first (non-trivial) all-multiplicity expressions for scattering amplitudes on any strong background. Their derivation brings together a rich set of geometric techniques which will be described in~\cite{Adamo:2020tba}, but the results presented here emulate momentum space expressions built from familiar ingredients. The distinction from tree-level MHV scattering on a trivial background lies in the breaking of 4-momentum conservation by the background (captured by the gauge and gravitational Volkov exponents) and (in the gravitational case) explicit tail terms from scattering off the SDPW background.

Several features of our formulae are surprising. Firstly, it is not apparent how they arise from tools familiar to the mainstream scattering amplitudes community. The presence of the free function $f(x^-)$ describing the SDPW background means that standard locality and unitarity arguments cannot be used to constrain the structure of tree amplitudes. Recently, it was shown in QED that gauge invariance could provide analytic constraints on amplitudes in strong plane waves~\cite{Ilderton:2020rgk}; we hope that further developments in this vein could enable robust independent checks (or alternative derivations) of our formulae. The fact that both formulae contain only a \emph{single} lightfront integral at arbitrary multiplicity is remarkable, but explained by their origin in a twistorial expansion of local space-time generating functionals~\cite{Mason:2005zm,Mansfield:2005yd,Mason:2008jy,Adamo:2013tja,Adamo:2020tba}.

There are several directions for future research suggested by this work. The first is to extend away from the MHV sector to other helicity configurations. This topic is under active investigation, and we will present conjectural N$^k$MHV expressions at arbitrary multiplicity in~\cite{Adamo:2020tba}. Another is to extend to Einstein-Yang-Mills amplitudes, extending the formulae of \cite{Adamo:2015gia,Roehrig:2017wvh} to incorporate self-dual backgrounds of both types.
More difficult is to relax the self-duality assumption on the background, considering general sandwich plane wave backgrounds. This is challenging, since the classical integrability and twistor theory underlying the derivation of the formulae here are lost, but it may be possible to use non-chiral ambitwistor string theory~\cite{Mason:2013sva} -- which encodes the full non-linearity of a generic strong background~\cite{Adamo:2014wea,Adamo:2018hzd} -- to make progress.

Colour-kinematics duality and double copy between gauge theory and gravity on strong backgrounds can be explored with these formulae. It should now be possible to test the proposals of~\cite{Adamo:2017nia,Adamo:2018mpq} at arbitrary multiplicity, as well as investigate the kinematic algebra associated to the SD sector~\cite{Monteiro:2011pc} in the presence of SD background fields.

Finally, one might imagine that \eqref{glMHV1} and \eqref{grMHV} are purely `academic,' as the SDPW background is complex; plane wave backgrounds relevant for phenomenological strong field QED, QCD or gravity are real. However, it has been shown that certain back-reaction effects (e.g., beam depletion~\cite{Seipt:2016fyu}) in strong field QFT can be computed through perturbation theory around fixed \emph{complex} background fields~\cite{Ilderton:2017xbj}. Further, the intrinsic chirality of introducing spin via the Newman-Janis trick~\cite{Newman:1965tw} indicates potential applications for these methods in gravitational wave physics. 

\medskip

\textit{Acknowledgments:} We thank Lance Dixon, Maciej Dunajski, Anton Ilderton and Alexander MacLeod for interesting discussions and comments. TA is supported by a Royal Society University Research Fellowship.  AS is supported by a Mathematical Institute Studentship, Oxford.

\bibliographystyle{JHEP}
\bibliography{sdpw1}

\end{document}